\documentclass[a4paper,12pt]{article}
\usepackage{amsfonts,amsthm,amsmath,amssymb,graphicx,hyperref,color,youngtab}

\newcommand{\be}{\begin{equation}}
\newcommand{\bea}{\begin{eqnarray}}

\newcommand{\ee}{\end{equation}}
\newcommand{\eea}{\end{eqnarray}}

\begin{document}

\makeatletter
\@addtoreset{equation}{section}
\makeatother
\renewcommand{\theequation}{\thesection.\arabic{equation}}

\rightline{}
\vspace{1.8truecm}

\vspace{15pt}


{\LARGE{  
\centerline{\bf Eigenvalue Dynamics for Multimatrix Models} 
}}  

\vskip.5cm 

\thispagestyle{empty}
    {\large \bf 
\centerline{Robert de Mello Koch\footnote{ {\tt robert@neo.phys.wits.ac.za}}, 
David Gossman\footnote{ {\tt dmgossman@gmail.com}}, 
Lwazi Nkumane\footnote{ {\tt lwazi.nkumane@gmail.com}}}
\centerline{ and   Laila Tribelhorn\footnote{ {\tt laila.tribelhorn@gmail.com}}}}

\vspace{.4cm}
\centerline{{\it National Institute for Theoretical Physics ,}}
\centerline{{\it School of Physics and Mandelstam Institute for Theoretical Physics,}}
\centerline{{\it University of Witwatersrand, Wits, 2050, } }
\centerline{{\it South Africa } }

\vspace{1.4truecm}

\thispagestyle{empty}

\centerline{\bf ABSTRACT}

\vskip.4cm 
\noindent By performing explicit computations of correlation functions, we find evidence that there is a sector of the two matrix model
defined by the $SU(2)$ sector of ${\cal N}=4$ super Yang-Mills theory, that can be reduced to eigenvalue dynamics.
There is an interesting generalization of the usual Van der Monde determinant that plays a role.
The observables we study are the BPS operators of the $SU(2)$ sector and include traces of products of both matrices, 
which are genuine multi matrix observables.
These operators are associated to supergravity solutions of string theory.

\setcounter{page}{0}
\setcounter{tocdepth}{2}

\newpage

\tableofcontents

\setcounter{footnote}{0}

\linespread{1.1}
\parskip 4pt

{}~
{}~

\section{Motivation}

The large $N$ expansion continues to be a promising approach towards the strong coupling dynamics of quantum
field theories. 
For example, 't Hooft's proposal that the large $N$ expansions of Yang-Mills theories are equivalent to the usual 
perturbation expansion in terms of topologies of worldsheets in string theory\cite{'tHooft:1973jz} has been realized 
concretely in the AdS/CFT correspondence\cite{Maldacena:1997re}.
Besides the usual planar limit where classical operator dimensions are held fixed as we take $N\to\infty$, there are non-planar
large $N$ limits of the theory \cite{Balasubramanian:2001nh} defined by considering operators with a bare dimension 
that is allowed to scale with $N$ as we take $N\to\infty$.
These limits are also relevant for the AdS/CFT correspondence.
Indeed, operators with a dimension that scales as $N$ include operators relevant for the description of 
giant graviton branes\cite{McGreevy:2000cw,Hashimoto:2000zp,Grisaru:2000zn} while operators with a dimension of 
order $N^2$ include operators that correspond to new geometries 
in supergravity\cite{Lin:2004nb,Corley:2001zk,Berenstein:2004kk}.
Despite these convincing motivations carrying out the large $N$ expansion for most matrix models is still beyond our 
current capabilities.
\\
\\
One class of models for which the large $N$ expansion can be computed are the singlet sector of matrix quantum mechanics of a 
single hermitian matrix\cite{BIPZ}.
We can also consider a complex matrix model as long as we restrict ourselves to potentials that are analytic in $Z$ (summed with the 
dagger of this which needs to be added to get a real potential) and observables constructed out of traces of
a product of $Z$s or out of a product of $Z^\dagger$s\cite{Ginibre:1965zz}.
In these situations we can reduce the problem to eigenvalue dynamics.
This is a huge reduction in degrees of freedom since we have reduced from $O(N^2)$ degrees of freedom, associated to the
matrix itself, to $O(N)$ eigenvalue degrees of freedom.
Studying saddle points of the original matrix action does not reproduce the large $N$ values of observables.
This is a consequence of the large number of degrees of freedom: we expect fluctuations to be suppressed by $1/N^2$
so that if $N^2$ variables in total are fluctuating, then we can have fluctuations of size $1/N^2\,\times\, N^2\sim 1$ which are not 
suppressed as $N\to\infty$.
In terms of eigenvalues there are only $N$ variables fluctuating so that fluctuations are bounded by $N\,\times\, 1/N^2\sim 1/N$
which vanishes as $N\to\infty$.
Thus, classical eigenvalue dynamics captures the large $N$ limit.
For example, one can formulate the physics of the planar limit by using the density of eigenvalues as a dynamical variable.
The resulting collective field theory defines a field theory that explicitly has $1/N$ as a coupling 
constant\cite{Jevicki:1979mb,Jevicki:1980zg}.
It has found both application in the context of the $c=1$ string\cite{Demeterfi:1991tz,Demeterfi:1991nw,Demeterfi:1991cw} 
and in descriptions of the LLM geometries\cite{Donos:2005vm}.
\\
\\
Standard arguments show that eigenvalue dynamics corresponds to a familiar system: non-interacting fermions in an external
potential\cite{BIPZ}.
This makes the description extremely convenient because the fermion dynamics is rather simple.
This eigenvalue dynamics is also a very natural description of the large $N$ but non-planar limits discussed above.
Giant graviton branes which have expanded into the $AdS_5$ of the spacetime correspond to highly excited fermions
or, equivalently, to single highly excited eigenvalues: the giant graviton is an eigenvalue\cite{Hashimoto:2000zp,Berenstein:2004kk}.
Giant graviton branes which have expanded into the $S^5$ of the spacetime correspond to holes in the Fermi sea, and
hence to collective excitations of the eigenvalues where many eigenvalues are excited\cite{Berenstein:2004kk}.
Half-BPS geometries also have a natural interpretation in terms of the eigenvalue dynamics: every fermion state
can be identified with a particular supergravity geometry\cite{Corley:2001zk,Berenstein:2004kk}.
The map between the two descriptions was discovered by Lin, Lunin and Maldacena in \cite{Lin:2004nb}.
The fermion state can be specified by stating which states in phase space are occupied by a fermion, so we can divide phase space up
into occupied and unoccupied states.
By requiring regularity of the corresponding supergravity solution exactly the same structure arises: the complete set of regular
solutions are specified by boundary conditions obtained by dividing a certain plane into black (identified with occupied states in 
the fermion phase space) and white (unoccupied states) regions. See \cite{Lin:2004nb} for the details.
\\
\\
Our main goal in this paper is to ask if a similar eigenvalue description can be constructed for a two matrix model.
Further, if such a construction exists, does it have a natural AdS/CFT interpretation?
Work with a similar motivation but focusing on a different set of questions has appeared 
in\cite{Masuku:2009qf,Kimura:2009ur,Masuku:2011pm,Masuku:2014wxa,Masuku:2015vta}.
We will consider the dynamics of two complex matrices, corresponding to the $SU(2)$ sector of ${\cal N}=4$ super Yang-Mills theory.
Further we consider the theory on $R\times S^3$ and expand all fields in spherical harmonics of the $S^3$.
We will consider only the lowest $s$-wave components of these expansions so that the matrices are constant on the $S^3$.
The reduction to the $s$-wave will be motivated below.
In this way we find a matrix model quantum mechanics of two complex matrices.
Expectation values are computed as follows
\bea
\langle\cdots\rangle =\int [dZ dZ^\dagger dY dY^\dagger] e^{-S}\cdots \label{mmvevs}
\eea
At first sight it appears that any attempts to reduce (\ref{mmvevs}) to an eigenvalue description are doomed to fail: the integral
in (\ref{mmvevs}) runs over two independent complex matrices $Z$ and $Y$ which will almost never be simultaneously 
diagonalizable.
However, perhaps there is a class of questions, generalizing the singlet sector of a single hermitian matrix model, that can
be studied using eigenvalue dynamics.
To explore this possibility, let's review the arguments that lead to eigenvalue dynamics for a single complex matrix $Z$.
We can use the Schur decomposition\cite{Ginibre:1965zz,Kristjansen:2002bb,Takayama:2005yq},
\bea
   Z=U^\dagger D U\label{SchurDecomp}
\eea
with $U$ a unitary matrix and $D$ is an upper triangular matrix, to explicitly change variables. 
Since we only consider observables that depend on the eigenvalues (the diagonal elements of $D$)
we can integrate $U$ and the off diagonal elements of $D$ out of the model, leaving only the eigenvalues.
The result of the integrations over $U$ and the off diagonal elements of $D$ is a non trivial Jacobian.
Denoting the eigenvalues of $Z$ by $z_i$, those of $Z^\dagger$ are given by complex conjugation, $\bar z_i$.
The resulting Jacobian is\cite{Ginibre:1965zz}
\bea
   J=\Delta (z)\Delta (\bar z) 
\eea
where
\bea
  \Delta (z)&=&\left|
\begin{matrix}
1     &1      &\cdots &1\cr
z_1 &z_2  &\cdots &z_N\cr
\vdots &\vdots &\vdots\,\,\vdots\,\,\vdots &\vdots\cr
z_1^{N-1} &z_2^{N-1} &\cdots &z_N^{N-1}
\end{matrix}
\right|\cr
&=&\prod_{j>k}^N (z_j-z_k)
\eea
is the usual Van der Monde determinant.
A standard argument now maps this into non-interacting fermion dynamics\cite{BIPZ}.
Trying to apply a very direct change of variables argument to the two matrix model problem appears difficult.
There is however an approach which both agrees with the above non-interacting fermion dynamics and can be generalized
to the two matrix model.
The idea is to construct a basis of operators that diagonalizes the inner product of the free theory.
The construction of an orthogonal basis, given by the Schur polynomials, was achieved in \cite{Corley:2001zk}.
Each Schur polynomial $\chi_R (Z)$ is labeled by a Young diagram $R$ with no more than $N$ rows.
In \cite{Corley:2001zk} the exact (to all order in $1/N$) two point function of Schur polynomials was constructed.
The result is
\bea
\langle\chi_R(Z)\chi_S(Z^\dagger)\rangle = f_R\delta_{RS}\label{schur2pnt}
\eea
where all spacetime dependence in the correlator has been suppressed.
This dependence is trivial as it is completely determined by conformal invariance.
The notation $f_R$ denotes the product of the factors of Young diagram $R$.
Remarkably there is an immediate and direct connection to non-interacting fermions: the fermion wave function can be written as
\bea
  \psi_R (\{z_i,\bar z_i\})=\chi_R(Z)\Delta (z)e^{-{1\over 2}\sum_i z_i\bar z_i}\label{swf}
\eea 
This relation can be understood as a combination of the state operator correspondence (we associate a Schur polynomial
operator on $R^{4}$ to a wave function on $R\times S^3$) and the reduction to eigenvalues (which is responsible for the
$\Delta (z)$ factor)\cite{Berenstein:2004kk}.
In this map the number of boxes in each row of $R$ determines the amount by which each fermion is excited.
In this way, each row in the Young diagram corresponds to a fermion and hence to an eigenvalue.
Having one very long row corresponds to exciting a single fermion by a large amount, which corresponds to a single
large (highly excited) eigenvalue. 
In the dual AdS gravity, a single long row is a giant graviton brane that has expanded in the AdS$_5$ space.
Having one very long column corresponds to exciting many fermions by a single quantum, which corresponds to many
eigenvalues excited by a small amount. 
In the dual AdS gravity, a single long column is a giant graviton brane that has expanded in the $S^5$ space.
\\
\\
The first questions we should tackle when approaching the two matrix problem should involve operators built using many
$Z$ fields and only a few $Y$ fields.
In this case at least a rough outline of the one matrix physics should be visible, and experience with the one matrix model
will prove to be valuable.

\begin{figure}[h]
        \centering
                \includegraphics[width=0.4\textwidth]{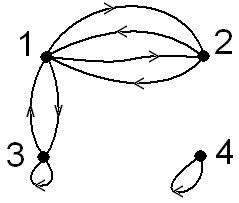}
        \label{dilpicjpg}
        \caption{An example of a graph labeling an operator with a definite scaling dimension. 
                         Each node corresponds to an eigenvalue.
                         Edges connect the different nodes so that the eigenvalues are interacting.}
\end{figure}

\noindent For the case of two matrices we can again construct a basis of operators that again diagonalizes the free field two point function.
These operators $\chi_{R,(r,s)ab}(Z,Y)$ are a generalization of the Schur polynomials, called restricted Schur 
polynomials\cite{Balasubramanian:2004nb,Bhattacharyya:2008rb,Bhattacharyya:2008xy}.
They are labeled by three Young diagrams ($R,r,s$) and two multiplicity labels ($a,b$).
For an operator constructed using $n$ $Z$s and $m$ $Y$s, $R\vdash n+m$, $r\vdash n$ and $s\vdash m$.
The multiplicity labels distinguish between different copies of the $(r,s)$ irreducible representation of $S_n\times S_m$ that arise
when we restrict the irreducible representation $R$ of $S_{n+m}$ to the $S_n\times S_m$ subgroup.
The two point function is
\bea
\langle \chi_{R,(r,s)ab}(Z,Y)\chi_{T,(t,u)cd}(Z^\dagger ,Y^\dagger)\rangle
=f_R{{\rm hooks}_R\over {\rm hooks}_r{\rm hooks}_s}\delta_{RT}\delta_{rt}\delta_{su}\delta_{ac}\delta_{bd}
\eea
where $f_R$ was defined after (\ref{schur2pnt}) and ${\rm hooks}_a$ denotes the product of the hook lengths associated
to Young diagram $a$.
These operators do not have a definite dimension.
However, they only mix weakly under the action of the dilatation operator and they form a convenient basis in which
to study the spectrum of anomalous dimensions\cite{DeComarmond:2010ie}.
This action has been diagonalized in a limit in which $R$ has order 1 rows (or columns), $m\ll n$ and $n$ is of order $N$.
Operators of a definite dimension are labeled by graphs composed of nodes that are traversed by oriented 
edges\cite{Koch:2011hb,deMelloKoch:2012ck}.
There is one node for each row, so that each node corresponds to an eigenvalue.
The directed edges start and end on the nodes.
There is one edge for each $Y$ field and the number of oriented edges ending on a node must equal the number of
oriented edges emanating from a node.
See figure 1 for an example of a graph labeling an operator.
This picture, derived in the Yang-Mills theory, has an immediate and compelling interpretation in the dual gravity:
each node corresponds to a giant graviton brane and the directed edges are open string excitations of these branes.
The constraint that the number of edges ending on a node equals the number of edges emanating from the node is simply
encoding the Gauss law on the brane world volume, which is topologically an $S^3$.
For this reason the graphs labeling the operators are called Gauss graphs.
If we are to obtain a system of non-interacting eigenvalues, we should only consider Gauss graphs that have no directed edges
stretching between nodes.
See figure 2 for an example.
In fact, these all correspond to BPS operators.
We thus arrive at a very concrete proposal:

\begin{figure}[hb]
        \centering
                \includegraphics[width=0.3\textwidth]{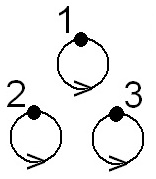}
        \label{configsjpg}
        \caption{An example of a graph labeling a BPS operator. 
                         Each node corresponds to an eigenvalue.
                         There are no edges connecting the different nodes so that these eigenvalues are not interacting.}
\end{figure}

{\vskip 0.5cm}

\noindent{\bf
If there is a free fermion description arising from the eigenvalue dynamics of the two matrix model, it will describe the BPS
operators of the $SU(2)$ sector.}

{\vskip 0.5cm}

\noindent The BPS operators are associated to supergravity solutions of string theory.
Indeed, the only one-particle states saturating the BPS bound in gravity are associated to massless particles
and lie in the supergravity multiplet.
Thus, eigenvalue dynamics will reproduce the supergravity dynamics of the gravity dual.
\\
\\
The BPS operators are all constructed from the $s$-wave of the spherical harmonic expansion on $S^3$\cite{Berenstein:2004kk}.
This is our motivation for only considering operators constructed using the $s$-wave of the fields $Y$ and $Z$.
One further comment is that it is usually not consistent to simply restrict to a subset of the dynamical degrees of freedom.
Indeed, this is only possible if the subset of degrees of freedom dynamically decouples from the rest of the theory.
In the case that we are considering this is guaranteed to be the case, in the large $N$ limit, because the Chan-Paton indices 
of the directed edges are frozen at large $N$ \cite{Koch:2011hb}.  
\\
\\
We should mention that eigenvalue dynamics as dual to supergravity has also been advocated by Berenstein and his
collaborators\cite{Berenstein:2005aa,Berenstein:2006yy,Berenstein:2007wz,Berenstein:2008jn,Berenstein:2008me,Berenstein:2010dg,Berenstein:2014pma}.
See also \cite{Balasubramanian:2005mg,Vazquez:2006id,Chen:2007gh,Koch:2008ah} for related studies.
Using a combination of numerical and physical arguments, which are rather different to the route we have followed, 
compelling evidence for this proposal has already been found.
The basic idea is that at strong coupling the commutator squared term in the action forces the Higgs fields
to commute and hence, at strong coupling, the Higgs fields of the theory should be simultaneously diagonalizable.
In this case, an eigenvalue description is possible.
Notice that our argument is a weak coupling argument, based on diagonalization of the one loop dilatation operator, that 
comes to precisely the same conclusion.
In this article we will make some exact analytic statements that agree with and, in our opinion, refine some of
the physical picture of the above studies.
For example, we will start to make precise statements about what eigenvalue dynamics does and does not correctly reproduce.

\section{Eigenvalue Dynamics for AdS$_5\times$S$^5$}\label{gs}

To motivate our proposal for eigenvalue dynamics, we will review the ${1\over 2}$-BPS sector stressing the logic
that we will subsequently use.
The way in which a direct change of variables is used to derive the eigenvalue dynamics can be motivated by considering
a correlation function of some arbitrary observables $\cdots$ that are functions only of the eigenvalues.
Because we are considering BPS operators, correlators computed in the free field theory agree with the same computations
at strong coupling\cite{Lee:1998bxa}, so that we now work in the free field theory.
Performing the change of variables we find
\bea
\langle\cdots\rangle &=& \int [ dZ dZ^\dagger ] e^{-{\rm Tr}ZZ^\dagger}\cdots\cr
&=&\int\prod_{i=1}^N dz_id\bar z_i e^{-\sum_k z_z\bar z_k}\Delta (z)\Delta (\bar z)\cdots\cr
&=&\int\prod_{i=1}^N dz_id\bar z_i |\psi_{\rm gs}(\{z_i,\bar z_i\})|^2\cdots\nonumber
\eea
where the groundstate wave function is given by
\bea
  \psi_{\rm gs} (\{z_i,\bar z_i\})=\Delta (z)e^{-{1\over 2}\sum_i z_i\bar z_i}\label{gswf}
\eea 
We will shortly qualify the adjective ``groundstate''.
Under the state-operator correspondence, this wave function is the state corresponding to the identity operator. 
The above transformation is equivalent to the identification
\bea
  [dZ]e^{-{1\over 2}{\rm Tr}(ZZ^\dagger )}\leftrightarrow c \prod_{i=1}^N dz_i\, \psi_{\rm gs}(\{z_i,\bar z_i\})\label{replaceone}
\eea
where $c$ is a constant that arises from integrating over $U,U^\dagger$ and the off diagonal elements of $D$ in 
(\ref{SchurDecomp}).
The role of each of the elements of the wave function is now clear:
\begin{itemize}
\item[1.] Under the state operator correspondence, dimensions of operators map to energies of states. The dimensions of BPS
               operators are not corrected, i.e. they take their free field values. This implies an evenly spaced spectrum and hence 
               a harmonic oscillator wave function. This explains the $e^{-{1\over 2}\sum_i z_i\bar z_i}$ factor. It also suggests
              that the wavefunction will be a polynomial times this Gaussian factor.
\item[2.] There is a gauge symmetry $Z\to UZU^\dagger$ that is able to permute the eigenvalues. 
               Consequently we are discussing identical particles.
               Two matrices drawn at random from the complex Gaussian ensemble will not have degenerate eigenvalues,
               so we choose the particles to be fermions.
               This matches the fact that the wave function is a Slater determinant.
\item[3.] Under the transformation $Z\to e^{i\theta} Z$, $dZ$ transforms with charge $N^2$. Since $\prod_i dz_i$ has charge $N$,
               $c \psi_{\rm gs}(\{z_i,\bar z_i\})$ must have charge $N(N-1)$.
               The constant $c$ is obtained by integrating over the off diagonal elements of $D$ in (\ref{SchurDecomp}).
               Thus, $c$ has charge ${1\over 2}N(N-1)$ and $\psi_{\rm gs}(\{z_i,\bar z_i\})$ itself has the same charge\footnote{We
               are assuming that any non-trivial measure depends only on the eigenvalues. This is a guess and we do not know a proof of
               this. We will make this assumption for the two matrix model as well.}.
\item[4.] If we assign the dimension $\big[Z\big]=L$ it is clear that both $\psi_{\rm gs}(\{z_i,\bar z_i\})$ and $c$ must have
               dimension ${1\over 2}N(N-1)$. 
\end{itemize}
The wave function (\ref{gswf}) satisfies these properties.
Further, if we require that the wavefunction is a polynomial in the eigenvalues $z_i$ times the exponential 
$e^{-{1\over 2}\sum_i z_i\bar z_i}$, then (\ref{gswf}) is the state of lowest energy (we did not write down a Hamiltonian, but
any other wave function has more nodes and hence a higher energy) so it deserves to be called the ground state.
The wave function (\ref{gswf}) is the state corresponding to the AdS$_5\times$S$^5$ spacetime in the ${1\over 2}$-BPS sector.
\\
\\
The above discussion can be generalized to write down a wave function corresponding to the AdS$_5\times$S$^5$ spacetime in the
$SU(2)$ sector.
The equation (\ref{replaceone}) is generalized to
\bea
  [dZ dY]e^{-{1\over 2}{\rm Tr}(ZZ^\dagger )-{1\over 2}{\rm Tr}(YY^\dagger )}\to 
c \prod_{i=1}^N dz_i dy_i\, \Psi_{\rm gs}(\{z_i,\bar z_i,y_i,\bar y_i\})
\eea
where $c$ is again a constant coming from integrating the non-eigenvalue variables out. 
The wave function must obey the following properties:
\begin{itemize}
\item[1.] Our wave functions again describe states that correspond to BPS operators. The dimensions of the BPS operators
               take their free field values, implying an evenly spaced spectrum and hence a harmonic oscillator wave function. 
               This suggests the wave function is a polynomial times the Gaussian factor
               $e^{-{1\over 2}\sum_i z_i\bar z_i-{1\over 2}\sum_i y_i\bar y_i}$ factor.
\item[2.] There is a gauge symmetry $Z\to UZU^\dagger$ and $Y\to UYU^\dagger$ that is able to permute the eigenvalues. 
               Consequently we are discussing $N$ identical particles.
               Matrices drawn at random will not have degenerate eigenvalues, so we choose the particles to be fermions.
               Thus we expect the wave function is a Slater determinant.
\item[3.] Under the transformation $Z\to e^{i\theta} Z$ and $Y\to Y$ the measure $dZ dY$ transforms with charge $N^2$. 
               Since $\prod_i dz_i dy_i$ has charge $N$ and $c$ has charge ${1\over 2}N(N-1)$, the wave function
               $\Psi_{\rm gs}(\{z_i,\bar z_i,y_i,\bar y_i\})$  must have charge ${1\over 2}N(N-1)$.
               Similarly, under the transformation $Z\to Z$ and $Y\to e^{i\theta}Y$ the measure $dZ dY$ transforms with charge $N^2$. 
               Since $\prod_i dz_i dy_i$ has charge $N$ and again $c$ has charge ${1\over 2}N(N-1)$, the wave function
              $\Psi_{\rm gs}(\{z_i,\bar z_i,y_i,\bar y_i\})$   should have charge ${1\over 2} N(N-1)$.
\item[4.] If we assign the dimension $\big[Z\big]=L=\big[Y\big]$ it is clear that both $\Psi_{\rm gs}(\{z_i,\bar z_i,y_i,\bar y_i\})$ and
               $c$ must have dimension $N(N-1)$.
\item[5.] The probability density associated to a single particle $\rho_{\rm gs}(z_1,\bar z_1,y_1,\bar y_1)$ must have an
               $SO(4)$ symmetry,  i.e. it should be a function of $|z_i|^2+|y_i|^2$.
\end{itemize}
The single particle probability density referred to in point 5 above is given, for any state $\Psi (\{z_i,\bar z_i,y_i,\bar y_i\})$ 
as usual, by
\begin{equation}
\rho(z_1,\bar z_1,y_1,\bar y_1)=\int \prod_{i=2}^N dz_id\bar z_i dy_id\bar y_i|\Psi (\{z_i,\bar z_i,y_i,\bar y_i\})|^2
\label{MMsurf}
\end{equation}
There is a good reason why the single particle probability density is an interesting quantity to look at: at short distances the
eigenvalues feel a repulsion from the Slater determinant, which vanishes when two eigenvalues are equal.
At long distances the confining harmonic oscillator potential dominates, ensuring the eigenvalues are clumped together in some
finite region and do not wander off to infinity.
In the end we expect that at large $N$ the locus where the eigenvalues lie defines a specific surface, generalizing the idea
of a density of eigenvalues for the single matrix model.
This large $N$ surface is captured by $\rho(z_1,\bar z_1,y_1,\bar y_1)$.
We will make this connection more explicit in a later section.
\\
\\
There appears to be a unique wave function singled out by the above requirements.
It is given by
\bea
\Psi_{\rm gs} (\{z_i,\bar z_i,y_i,\bar y_i\}) = {\cal N}\Delta (z,y)e^{-{1\over 2}\sum_k z_z\bar z_k -{1\over 2}\sum_k y_z\bar y_k}
\eea
where
\bea
\Delta (z,y)&=&\left|
\begin{matrix}
y_1^{N-1}     &y_2^{N-1}      &\cdots &y_N^{N-1}\cr
z_1y_1^{N-2} &z_2y_2^{N-2}  &\cdots &z_Ny_N^{N-2}\cr
\vdots &\vdots &\vdots\,\,\vdots\,\,\vdots &\vdots\cr
z_1^{N-2}y_1 &z_2^{N-2}y_2 &\cdots &z_N^{N-2}y_N\cr
z_1^{N-1} &z_2^{N-1} &\cdots &z_N^{N-1}
\end{matrix}
\right|\cr
&=&\prod_{j>k}^N (z_jy_k-y_jz_k)
\eea
generalizes the usual Van der Monde determinant and ${\cal N}$ is fixed by normalizing the wave function.
Normalizing the wave function in the state picture corresponds to choosing a normalization in the original matrix model
so that the expectation value of $1$ is $1$.
\\
\\
We can provide detailed tests of this wave function by using the equation
\bea
\int [dY dZ dY^\dagger dZ^\dagger ] e^{-{\rm Tr}(ZZ^\dagger)-{\rm Tr}(YY^\dagger)} \cdots
=\int\prod_{i=1}^N dz_id\bar z_i dy_i d\bar y_i |\Psi_{\rm gs} (\{z_i,\bar z_i,y_i,\bar y_i\})|^2\cdots
\eea
to compute correlators of observables (denoted by $\cdots$ above) that depend only on the eigenvalues.
We have already argued above that we expect that these observables are the BPS operators of the CFT.
As a first example, consider correlators of traces $O_J={\rm Tr}(Z^J)$.
These can be computed exactly in the matrix model, using a variety of different techniques - see for example
\cite{Ginibre:1965zz,Corley:2002mj,Kristjansen:2002bb}.
The result is
\bea
\langle {\rm Tr}(Z^J){\rm Tr}(Z^{\dagger J})\rangle
={1\over J+1}\Big[{(J+N)!\over (N-1)!}-{N!\over (N-J-1)!}\Big]
\eea
if $J<N$ and
\bea
\langle {\rm Tr}(Z^J){\rm Tr}(Z^{\dagger J})\rangle
={1\over J+1} {(J+N)!\over (N-1)!}
\eea
if $J\ge N$.
These expressions could easily be expanded to generate the $1/N$ expansion if we wanted to do that.
We would now like to consider the eigenvalue computation.
It is useful to write the wave function as
\bea
\Psi_{\rm gs}(\{z_i,\bar z_i,y_i,\bar y_i\})={\pi^{-N}\over\sqrt{N!}}\epsilon^{a_1 a_2 \cdots a_n} 
{z_{a_1}^0 y_{a_1}^{N-1}\over \sqrt{0!(N-1)!}}\cdots {z_{a_k}^{k-1}y_{a_k}^{N-k}\over\sqrt{(k-1)!(N-k)!}} \cdots\cr
\cdots {z_{a_N}^{N-1}y_{a_N}^0\over\sqrt{(N-1)!0!}} 
e^{-{1\over 2}\sum_q z_q\bar z_q-{1\over 2}\sum_q y_q\bar y_q}
\eea
The gauge invariant observable in this case is given by
\bea
{\rm Tr}(Z^J){\rm Tr}(Z^{\dagger J})=\sum_{i=1}^N z_i\sum_{j=1}^N\bar z_j
\eea
It is now straightforward to find
\bea
\int\prod_{i=1}^N dz_id\bar z_i dy_i d\bar y_i |\Psi (\{z_i,\bar z_i,y_i,\bar y_i\})|^2 \sum_i z_i^J\sum_j\bar z_j^J
={1\over J+1} {(J+N)!\over (N-1)!}\label{eigentrace}
\eea
When evaluating the above integral, only the terms with $i=j$ contribute.
From this result we see that we have not reproduced traces with $J< N$ correctly - we don't even get the leading
large $N$ behavior right.
We have, however, correctly reproduced the exact answer (to all orders in $1/N$) of the two point function
for all single traces of dimension $N$ or greater.
For $J>N$ there are trace relations of the form
\bea
{\rm Tr}(Z^J)=\sum_{i,j,...,k}c_{ij...k}{\rm Tr}(Z^i){\rm Tr}(Z^j)\cdots {\rm Tr}(Z^k)
\eea
$i,j,...,k\le N$ and $i+j+\cdots+k=J$.
The fact that we reproduce two point correlators of traces with $J>N$ exactly implies that we also start to reproduce sums 
of products of traces of less than $N$ fields.
This suggests that the important thing is not the trace structure of the operator, but rather the dimension of the state.
\\
\\
The fact that we only reproduce observables that have a large enough dimension is not too surprising.
Indeed, supergravity can't be expected to correctly describe the back reaction of a single graviton or a single string.
To produce a state in the CFT dual to a geometry that is different from the AdS vacuum one needs to allow a number of 
AdS giant gravitons (eigenvalues) to condense.
The eigenvalue dynamics is correctly reproducing the two point function of traces when their energy is greater than
that required to blow up into an AdS giant graviton.
\\
\\
With a very simple extension of the above argument we can argue that we also correctly reproduce the correlator
$\langle {\rm Tr}(Y^J){\rm Tr}(Y^{\dagger J})\rangle$ with $J\ge N$.
A much more interesting class of observables to consider are mixed traces, which contain both $Y$ and $Z$ fields.
To build BPS operators using both $Y$ and $Z$ fields we need to construct symmetrized traces.
A very convenient way to perform this construction is as follows
\bea
   {\cal O}_{J,K}={J!\over (J+K)!} {\rm Tr}\left(Y{\partial\over\partial Z}\right)^K {\rm Tr}(Z^{J+K})
\eea
The normalization up front is just the inverse of the number of terms that appear.
With this normalization, the translation between the matrix model observable and an eigenvalue observable is
\bea 
   {\cal O}_{J,K} \leftrightarrow \sum_i z_i^J y_i^K\label{mysteryone}
\eea
Since we could not find this computation in the literature, we will now explain how to evaluate the matrix model two
point function exactly, in the free field theory limit.
Since the dimension of BPS operators are not corrected, this answer is in fact exact.
To start, perform the contraction over the $Y,Y^\dagger$ fields
\bea
\langle {\cal O}_{J,K}{\cal O}_{J,K}^\dagger\rangle &=&
\left({J!\over (J+K)!}\right)^2
\langle
{\rm Tr}\left(Y{\partial\over\partial Z}\right)^K {\rm Tr}(Z^{J+K})
{\rm Tr}\left(Y^\dagger {\partial\over\partial Z^\dagger }\right)^K {\rm Tr}(Z^{\dagger\, J+K})\rangle\cr
&=&
\left({J!\over (J+K)!}\right)^2 K!
\langle
{\rm Tr}\left({\partial\over\partial Z}{\partial\over\partial Z^\dagger }\right)^K 
{\rm Tr}(Z^{J+K}) {\rm Tr}(Z^{\dagger\, J+K})
\rangle
\eea
Given the form of the matrix model two point function
\bea
   \langle Z_{ij}Z^\dagger_{kl}\rangle =\delta_{il}\delta_{jk}
\eea
we know that we can write any free field theory correlator as
\bea
\langle\cdots\rangle =
  e^{{\rm Tr}\left({\partial\over\partial Z}{\partial\over\partial Z^\dagger }\right)}\cdots\Big|_{Z=Z^\dagger=0}
\eea
Using this identity we now find
\bea
\langle {\cal O}_{J,K}{\cal O}_{J,K}^\dagger\rangle 
= \left({J!\over (J+K)!}\right)^2 K! {(J+K)!\over J!}
\langle {\rm Tr}(Z^{J+K}) {\rm Tr}(Z^{\dagger\, J+K}) \rangle
\eea
Thus, the result of the matrix model computation is
\bea
\langle {\cal O}_{J,K}{\cal O}_{J,K}^\dagger\rangle =
{J!K!\over (J+K+1)!}\left[
{(J+K+N)!\over (N-1)!}-{N!\over (N-J-K-1)!}\right]
\eea
if $J+K<N$ and
\bea
\langle {\cal O}_{J,K}{\cal O}_{J,K}^\dagger\rangle =
{J!K!\over (J+K+1)!}{(J+K+N)!\over (N-1)!}
\eea
if $J+K \ge N$.
Notice that for these two matrix observables we again get a change in the form of the correlator as the dimension
of the trace passes $N$.

Next, consider the eigenvalue computation. We need to perform the integral
\bea
\langle {\cal O}_{J,K} {\cal O}_{J,K}^\dagger\rangle
=\int\prod_{i=1}^N dz_id\bar z_i dy_id\bar y_i |\Psi_{\rm gs}(\{z_i,\bar z_i,y_i,\bar y_i\})|^2
\sum_{k=1}^N z_k^J y_k^K \sum_{j=1}^N \bar z_j^J \bar y_j^K
\eea
After some straightforward manipulations we have
\bea
\langle {\cal O}_{J,K} {\cal O}_{J,K}^\dagger\rangle
=\pi^{-2N} \int\prod_{i=1}^N dz_id\bar z_i dy_id\bar y_i  
{|z_1|^{0}|y_1|^{2N-2}\over 0! (N-1)!}
\cdots {|z_k|^{2k-2}|y_k|^{2N-2k}\over (k-1)! (N-k)!}\cdots \cr
{|z_N|^{2N-2} |y_N|^{0}\over (N-1)! 0!}
\times e^{-\sum_q z_q\bar z_q-\sum_q y_q\bar y_q} 
\sum_{k,j=1}^N z_k^J y_k^K \bar z_j^J \bar y_j^K
\eea
Only terms with $k=j$ contribute so that
\bea
\langle {\cal O}_{J,K} {\cal O}_{J,K}^\dagger\rangle
=\sum_{k=1}^N {(N-k+K)!\over (N-k)!}{(J+k-1)!\over (k-1)!}
={K!J!\over (K+J+1)!}{(J+K+N)!\over (N-1)!}
\eea
Thus, we again correctly reproduce the exact (to all orders in $1/N$) answer for the two point function of single trace operators 
of dimension $N$ or greater.
\\
\\
It is also interesting to consider multi trace correlators.
We will start with the correlator between a double trace and a single trace and we will again start with the matrix model computation
\bea
\langle O_{J_1,K_1} O_{J_2,K_2}O_{J_1+J_2,K_1+K_2}^\dagger\rangle
={J_1!\over (J_1+K_1)!} {J_2!\over (J_2+K_2)!} {(J_1+J_2)!\over (J_1+K_1+J_2+K_2)!}\times\cr
\langle {\rm Tr}\left(Y{\partial\over\partial Z}\right)^{K_1}{\rm Tr}(Z^{J_1+K_1})
{\rm Tr}\left(Y{\partial\over\partial Z}\right)^{K_2}{\rm Tr}(Z^{J_2+K_2})
{\rm Tr}\left(Y^\dagger {\partial\over\partial Z^\dagger}\right)^{K_1+K_2}{\rm Tr}(Z^{\dagger J_1+K_1J_2+K_2})\rangle\cr
\eea
We could easily set $K_1=K_2=0$ and obtain traces involving only a single matrix.
Begin by contracting all $Y,Y^\dagger$ fields to obtain
\bea
\langle O_{J_1,K_1} O_{J_2,K_2}O_{J_1+J_2,K_1+K_2}^\dagger\rangle
={J_1!\over (J_1+K_1)!} {J_2!\over (J_2+K_2)!} {(J_1+J_2)!\over (J_1+K_1+J_2+K_2)!} (K_1+K_2)!\times\cr
\langle {\partial\over\partial Z_{i_1 j_1}}\cdots {\partial\over\partial Z_{i_{K_1} j_{K_1}}}{\rm Tr}(Z^{J_1+K_1})
{\partial\over\partial Z_{i_{K_1+1} j_{K_1+1}}}\cdots 
{\partial\over\partial Z_{i_{K_1+K_2} j_{K_1+K_2}}}{\rm Tr}(Z^{J_2+K_2})\cr 
{\partial\over\partial Z^\dagger_{j_1 i_1}}\cdots
{\partial\over\partial Z^\dagger_{j_{K_1+K_2} i_{K_1+K_2}}}{\rm Tr}(Z^{\dagger J_1+K_1+J_2+K_2})\rangle\cr
\eea
It is now useful to integrate by parts with respect to $Z^\dagger$, using the identity
\bea
\langle {\partial\over\partial Z_{ij}}f(Z)\, g(Z)\, {\partial\over\partial Z^\dagger_{ji}}h(Z^\dagger)\rangle =
n_f \langle f(Z)\, g(Z)\, h(Z^\dagger)\rangle\label{usefulid}
\eea
where $f(Z)$ is of degree $n_f$ in $Z$.
Repeatedly using this identity, we find
\bea
\langle O_{J_1,K_1} O_{J_2,K_2}O_{J_1+J_2,K_1+K_2}^\dagger\rangle
={J_1!\over (J_1+K_1)!} {J_2!\over (J_2+K_2)!} {(J_1+J_2)!\over (J_1+K_1+J_2+K_2)!} (K_1+K_2)!\times\cr
{(J_1+K_1)!\over J_1!} {(J_2+K_2)!\over J_2!}
\langle {\rm Tr}(Z^{J_1+K_1}) {\rm Tr}(Z^{J_2+K_2}) {\rm Tr}(Z^{\dagger J_1+K_1+J_2+K_2})\rangle\cr
= {(J_1+J_2)! (K_1+K_2)!\over (J_1+K_1+J_2+K_2)!} 
\langle {\rm Tr}(Z^{J_1+K_1}) {\rm Tr}(Z^{J_2+K_2}) {\rm Tr}(Z^{\dagger J_1+K_1+J_2+K_2})\rangle\cr
\eea
This last correlator is easily computed.
For example, if $J_1+K_1<N$ and $J_2+K_2<N$ we have
\bea
\langle O_{J_1,K_1} O_{J_2,K_2}O_{J_1+J_2,K_1+K_2}^\dagger\rangle
={(J_1+J_2)! (K_1+K_2)!\over (J_1+K_1+J_2+K_2+1)!}\Big[&&\!\!\!\!\!\!\!\!\!\!\!\!\!\!\!
{(J_1+K_1+J_2+K_2+N)!\over (N-1)!}\cr
+{N!\over (N-J_1-K_1-J_2-K_2-1)!}\!\!\!&-&\!\!\!{(N+J_1+K_1)!\over (N-J_2-K_2-1)!}\cr
&-&\!\!\!{(N+J_2+K_2)!\over (N-J_1-K_1-1)!}\Big]\cr
&&
\eea
and if $J_1+K_1\ge N$ and $J_2+K_2\ge N$ we have
\bea
\langle O_{J_1,K_1} O_{J_2,K_2}O_{J_1+J_2,K_1+K_2}^\dagger\rangle
={(J_1+J_2)! (K_1+K_2)!\over (J_1+K_1+J_2+K_2+1)!}
{(J_1+K_1+J_2+K_2+N)!\over (N-1)!}\cr
&&
\eea
It is a simple exercise to check that, in terms of eigenvalues, we have
\bea
\langle {\cal O}_{J_1,K_1}{\cal O}_{J_2,K_2} {\cal O}_{J_1+J_2,K_1+K_2}^\dagger\rangle
&=&\int\prod_{i=1}^N dz_id\bar z_i dy_id\bar y_i |\Psi_{\rm gs}(\{z_i,\bar z_i,y_i,\bar y_i\})|^2\cr
&&\quad\times\sum_{k=1}^N z_k^{J_1} y_k^{K_1}\sum_{l=1}^N z_l^{J_2} y_l^{K_2} 
\sum_{j=1}^N \bar z_j^{J_1+J_2} \bar y_j^{K_1+K_2}\cr
&=&{(J_1+J_2)! (K_1+K_2)!\over (J_1+K_1+J_2+K_2+1)!}
{(J_1+K_1+J_2+K_2+N)!\over (N-1)!}\cr
&&
\eea
so that once again we have reproduced the exact answer as long as the dimension of each trace is not less than $N$.
The agreement that we have observed for multi trace correlators continues as follows: as long as the
dimension of each trace is greater than $N-1$ the matrix model and the eigenvalue descriptions agree and both give
\bea
\langle O_{J_1,K_1} O_{J_2,K_2}\cdots O_{J_n,K_n} O_{J,K}^\dagger\rangle
={J! K!\over (J+K+1)!}{(J+K+N)!\over (N-1)!}\delta_{J_1+\cdots+J_n,J}\delta_{K_1+\cdots+K_n,K}\cr
\eea
for the exact value of this correlator.
We have limited our selves to a single daggered observable in the above expression for purely technical reasons: it is
only in this case that we can compute the matrix model correlator using the identity (\ref{usefulid}).
It would be interesting to develop analytic methods that allow more general computations.
\\
\\
Finally, we can also test multi trace correlators with a dimension of order $N^2$.
A particularly simple operator is the Schur polynomial labeled by a Young diagram $R$ with $N$ rows and $M$ columns.
For this $R$ we have
\bea
\chi_R(Z)=(\det Z)^M=z_1^M z_2^M\cdots z_N^M
\eea
\bea
\chi_R(Z^\dagger )=(\det Z^\dagger)^M=\bar z_1^M \bar z_2^M\cdots \bar z_N^M
\eea
The dual LLM geometry is labeled by an annulus boundary condition that has an inner radius of $\sqrt{M}$ and an
outer radius of $\sqrt{M+N}$.
The two point correlator of this Schur polynomial is
\bea
\langle \chi_R(Z) \chi_R(Z^\dagger)\rangle
&=&\int\prod_{i=1}^N dz_id\bar z_i dy_id\bar y_i
\chi_R(Z)\chi_R(Z^\dagger )|\Psi_{\rm gs}(\{z_i,\bar z_i,y_i,\bar y_i\})|^2\cr
&=&\pi^{-2N} \int\prod_{i=1}^N dz_id\bar z_i dy_id\bar y_i  
{|z_1|^{0+2M}|y_1|^{2N-2}\over 0! (N-1)!}
\cdots {|z_k|^{2k-2+2M}|y_k|^{2N-2k}\over (k-1)! (N-k)!}\cr
&\times&\cdots {|z_N|^{2N-2+2M} |y_N|^{0}\over (N-1)! 0!}
\times e^{-\sum_q z_q\bar z_q-\sum_q y_q\bar y_q} \cr
&=&\prod_{i=1}^N {(i-1+M)!\over (i-1)!}
\eea
which is again the exact answer for this correlator.
\\
\\
After this warm up example we will now make a few comments that are relevant for the general case.
The details are much more messy, so we will not manage to make very precise statements.
We have however included this discussion as it does provide a guide as to when eigenvalue dynamics
is applicable.
A Schur polynomial labeled with a Young diagram $R$ that has row lengths $r_i$ is given in terms of eigenvalues as
(our labeling of the rows is defined by $r_1\ge r_2\ge\cdots\ge r_N$)
\bea
   \chi_R(Z)=
{\epsilon_{a_1 a_2 \cdots a_N} z_{a_1}^{N-1+r_1}z_{a_2}^{N-2+r_2}\cdots z_{a_{N}}^{r_{N}}\over
 \epsilon_{b_1 b_2 \cdots b_N} z_{b_1}^{N-1}z_{b_2}^{N-2}\cdots z_{b_{N-1}}} 
\eea
Using this expression, we can easily write the exact two point function as follows
\bea
\langle \chi_R(Z)\chi_R(Z^\dagger)\rangle
&=&{1\over N! \pi^{N}}
\int\prod_{i=1}^N dz_i d\bar z_i
\epsilon_{a_1 a_2 \cdots a_N} z_{a_1}^{N-1+r_1}z_{a_2}^{N-2+r_2}\cdots z_{a_{N}}^{r_{N}}\cr
&&\qquad\times\epsilon_{b_1 b_2 \cdots b_N} \bar z_{b_1}^{N-1+r_1}\bar z_{b_2}^{N-2+r_2}\cdots \bar z_{b_{N}}^{r_{N}}
e^{-\sum_k z_k\bar z_k}\cr
&=&\prod_{j=0}^{N-1}{(j+r_{N-j})!\over j!}=f_R\label{exactschur}
\eea
Using our wave function we can compute the two point function of Schur polynomials.
The result is
\bea
\langle \chi_R(Z) \chi_R(Z^\dagger)\rangle
&=&\int\prod_{i=1}^N dz_id\bar z_i dy_id\bar y_i
\chi_R(Z)\chi_R(Z^\dagger )|\Psi_{\rm gs}(\{z_i,\bar z_i,y_i,\bar y_i\})|^2\cr
&=&{1\over N! \pi^{N}}
\int\prod_{i=1}^N dz_i d\bar z_i
\epsilon_{a_1 a_2 \cdots a_N} |z_{a_1}|^{2N-2}|z_{a_2}|^{2N-4}\cdots |z_{a_{N-1}}|^{2}\cr
&&\qquad\times
{\epsilon_{b_1 b_2 \cdots b_N} z_{b_1}^{N-1+r_1}z_{b_2}^{N-2+r_2}\cdots z_{b_{N}}^{r_{N}}\over
\epsilon_{c_1 c_2 \cdots c_N} z_{c_1}^{N-1}z_{c_2}^{N-2}\cdots z_{a_{N-1}}}\cr
&&\qquad\times
{\epsilon_{d_1 d_2 \cdots d_N} \bar z_{d_1}^{N-1+r_1}\bar z_{d_2}^{N-2+r_2}\cdots \bar z_{d_{N}}^{r_{N}}\over
\epsilon_{e_1 e_2 \cdots e_N} \bar z_{e_1}^{N-1}\bar z_{e_2}^{N-2}\cdots \bar z_{e_{N-1}}}
e^{-\sum_k z_k\bar z_k}\label{evalschurtwopoint}
\eea
When the integration over the angles $\theta_i$ associated to $z_i=r_i e^{i\theta_i}$ are performed, a non-zero result
is only obtained if powers of the $z_i$ match the powers of the $\bar z_i$.
The difference between the above expression and the exact answer is simply that in the eigenvalue expression these powers
are separately set to be equal in the measure and in the product of Schur polynomials - there are two matchings, while in the 
exact answer the power of $z_i$ arising from the product of the measure and the product of Schur polynomials is matched to the power of $\bar z_i$  from the product of the measure and the product of Schur polynomials - there is a single matching happening.
Thus, the eigenvalue computation may miss some terms that are present in the exact answer\footnote{This is the reason why
(\ref{eigentrace}) only captures one of the terms present in the two point function for $J<N$.}.
For Young diagrams with a few corners and $O(N^2)$ boxes (the annulus above is a good example) the eigenvalues clump into
groupings, with each grouping collecting eigenvalues of a similar size corresponding to rows with a similar 
row length\cite{Koch:2008ah}.
This happens because the product of the Gaussian fall off $e^{-z\bar z}$ and a polynomial of fixed degree $|z^2|^n$ is
sharply peaked at $|z|=n$. 
Thus, for example if $r_i\approx M_1$ for $i=1,2,\cdots,{N\over 2}$ and $r_i\approx M_2$ for 
$i=1+{N\over 2},2+{N\over 2},\cdots,N$ with $M_1$ and $M_2$ well separated ($M_1-M_2\ge O(N)$), under the integral 
we can replace
\bea
{\epsilon_{b_1 b_2 \cdots b_N} z_{b_1}^{N-1+r_1}z_{b_2}^{N-2+r_2}\cdots z_{b_{N}}^{r_{N}}\over
\epsilon_{c_1 c_2 \cdots c_N} z_{c_1}^{N-1}z_{c_2}^{N-2}\cdots z_{a_{N-1}}}\to
\prod_{i=1}^{N\over 2} z_{a_i}^{M_1} z_{a_{i+{N\over 2}}}^{M_2}
\eea
After making a replacement of this type, we recover the exact answer.
This replacement is not exact - we need to appeal to large $N$ to justify it.
It would be very interesting to explore this point further and to quantify in general (if possible) what the corrections to the
above replacement are.
For Young diagrams with many corners, row lengths are not well separated and there is no similar grouping that occurs, so that the
eigenvalue description will not agree with the exact result, even at large $N$.
A good example of a geometry with many corners is the superstar\cite{Myers:2001aq}.
The corresponding LLM boundary condition is a number of very thin concentric annuli, so that we effectively obtain a gray disk,
signaling a singular supergravity geometry.
It is then perhaps not surprising that the eigenvalue dynamics does not correctly reproduce this two point correlator. 
\\
\\
Having discussed the two point function of Schur polynomials in detail, the product rule
\bea
   \chi_R(Z)\chi_S(Z)=\sum_T f_{RST}\chi_T(Z)
\eea
with $f_{RST}$ a Littlewood-Richardson coefficient, implies that there is no need to consider correlation functions of products of
Schur polynomials.

\section{Other backgrounds}\label{annulus}

In the ${1\over 2}$ BPS sector there is a wave function corresponding to every LLM geometry.
The (not normalized) wave function has already been given in (\ref{swf}).
In this section we consider the problem of writing eigenvalue wave functions that correspond to geometries
other than AdS$_5\times$S$^5$.
The simplest geometry we can consider is the annulus geometry considered in the previous section, where we argued
that the eigenvalue dynamics reproduces the exact correlator of the Schur polynomials dual to this geometry.
Our proposal for the state that corresponds to this LLM spacetime is
\bea
\Psi_{\rm LLM}(\{z_i,\bar z_i,y_i,\bar y_i\})={\pi^{-N}\over\sqrt{N!}}\epsilon^{a_1 a_2 \cdots a_n} 
{z_{a_1}^{M} y_{a_1}^{N-1}\over \sqrt{M!(N-1)!}}\cdots {z_{a_k}^{k-1+M}
y_{a_k}^{N-k}\over\sqrt{(k-1+M)!(N-k)!}} \cr
\cdots {z_{a_N}^{N-1+M}y_{a_N}^0\over\sqrt{(N-1+M)!0!}} 
e^{-{1\over 2}\sum_q z_q\bar z_q-{1\over 2}\sum_q y_q\bar y_q}
\eea
This is simply obtained by multiplying the ground state wave function by the relevant Schur polynomial and normalizing
the resulting state.
The connection between matrix model correlators and expectation values computed using the above wave function is
the following\footnote{The new normalization for matrix model correlators is needed to ensure that the identity operator has 
expectation value 1. This matches the normalization adopted in the eigenvalue description.}
\bea
\langle\,\,\cdots\,\,\rangle_{\rm LLM}&=&
{\langle \,\,\cdots \,\, \chi_R(Z)\chi_R(Z^\dagger)\rangle\over \langle\chi_R(Z)\chi_R(Z^\dagger)\rangle}\cr
&=&\int\prod_{i=1}^N dz_id\bar z_i dy_i d\bar y_i |\Psi_{\rm LLM} (\{z_i,\bar z_i,y_i,\bar y_i\})|^2\,\,\cdots
\eea

\noindent We can use this wave function to compute correlators that we are interested in.
Traces involving only $Z$s for example lead to
\bea
\langle {\rm Tr}(Z^J){\rm Tr}(Z^{\dagger J})\rangle_{\rm LLM}
&=&\int\prod_{i=1}^N dz_id\bar z_i dy_id\bar y_i |\Psi_{\rm LLM} (\{z_i,\bar z_i,y_i,\bar y_i\})|^2
\sum_{k=1}^N z_k^J\sum_{l=1}^N\bar z_l^J\cr
&=&\sum_{k=0}^{N-1}{(J+k+M)!\over (k+M)!}\cr
&=&{1\over J+1}\left[
{(J+M+N)!\over (M+N-1)!}-{(J+M)!\over (M-1)!}\right]
\eea
which agrees with the exact result, as long as $J>N-1$.
Thus, in this background, eigenvalue dynamics is correctly reproducing the same set of correlators as in the original
 AdS$_5\times$S$^5$ background.
Traces involving only $Y$ fields are also correctly reproduced
\bea
\langle {\rm Tr}(Y^J){\rm Tr}(Y^{\dagger J})\rangle_{\rm LLM}
=\int\prod_{i=1}^N dz_id\bar z_i dy_id\bar y_i |\Psi_{\rm LLM} (\{z_i,\bar z_i,y_i,\bar y_i\})|^2
\sum_{k=1}^N y_k^J\sum_{l=1}^N\bar y_l^J\cr
={1\over J+1} {(J+N)!\over (N-1)!}
\eea
where $J\ge N$.
Notice that these results are again exact, i.e. we reproduce the matrix model correlators to all orders in $1/N$.
Finally, let's consider the most interesting case of traces involving both matrices.
The LLM wave function we have proposed does not reproduce the exact matrix model computation.
The matrix model computation gives 
\bea
\langle {\cal O}_{J,K}{\cal O}_{J,K}^\dagger\rangle_{\rm LLM} &=&
\left({J!\over (J+K)!}\right)^2
\langle
{\rm Tr}\left(Y{\partial\over\partial Z}\right)^K {\rm Tr}(Z^{J+K})
{\rm Tr}\left(Y^\dagger {\partial\over\partial Z^\dagger }\right)^K {\rm Tr}(Z^{\dagger\, J+K})
\rangle_{\rm LLM}
\cr
&=&
\left({J!\over (J+K)!}\right)^2 K!
\langle
{\rm Tr}\left({\partial\over\partial Z}{\partial\over\partial Z^\dagger }\right)^K 
{\rm Tr}(Z^{J+K}) {\rm Tr}(Z^{\dagger\, J+K})
\rangle_{\rm LLM}\cr
&=& \left({J!\over (J+K)!}\right)^2 K! {(J+K)!\over J!}
\langle {\rm Tr}(Z^{J+K}) {\rm Tr}(Z^{\dagger\, J+K}) \rangle_{\rm LLM}\cr
&=&
{J!K!\over (J+K+1)!}\left[{(J+K+M+N)!\over (M+N-1)!}-{(J+K+M)!\over (M-1)!}\right]
\label{LLMmm}
\eea
if $J+K \ge N$.
Next, consider the eigenvalue computation. 
We need to perform the integral
\bea
\langle {\cal O}_{J,K} {\cal O}_{J,K}^\dagger\rangle_{\rm LLM,eigen}
&=&\int\prod_{i=1}^N dz_id\bar z_i dy_id\bar y_i |\Psi_{\rm LLM}(\{z_i,\bar z_i,y_i,\bar y_i\})|^2
\sum_{k=1}^N z_k^J y_k^K \sum_{j=1}^N \bar z_j^J \bar y_j^K\cr
&=&\sum_{k=1}^N {(N-k+K)!\over (N-k)!}{(J+M+k-1)!\over (M+k-1)!}\label{LLMeigen}
\eea
It is not completely trivial to compare (\ref{LLMmm}) and (\ref{LLMeigen}), but  it is already clear that they do not reproduce 
exactly the same answer.
To simplify the discussion, let's consider the case that $M=O(\sqrt{N})$.
In this case, in the large $N$ limit, we can drop the second term in (\ref{LLMmm}) to obtain
\bea
\langle {\cal O}_{J,K}{\cal O}_{J,K}^\dagger\rangle_{\rm LLM} =
{J!K!\over (J+K+1)!}{(J+K+M+N)!\over (M+N-1)!}(1+\cdots)
\label{appLLMmm}
\eea
where $\cdots$ stand for terms that vanish as $N\to\infty$.
In the sum appearing in (\ref{LLMeigen}), change variables from $k$ to $k'-M$ and again appeal to large $N$ to write
\bea
\langle {\cal O}_{J,K} {\cal O}_{J,K}^\dagger\rangle_{\rm LLM,eigen}
&=&\sum_{k'=M+1}^{M+N} {(N+M-k'+K)!\over (N+M-k')!}{(J+k'-1)!\over (k'-1)!}\cr
&=&\sum_{k'=1}^{M+N} {(N+M-k'+K)!\over (N+M-k')!}{(J+k'-1)!\over (k'-1)!}(1+\cdots)\cr
&=&{J!K!\over (J+K+1)!}{(J+K+M+N)!\over (M+N-1)!}(1+\cdots)\label{approxLLMeigen}
\eea
In the last two lines above $\cdots$ again stands for terms that vanish as $N\to\infty$.
Thus, we find agreement between (\ref{LLMmm}) and (\ref{LLMeigen}).
It is again convincing to see genuine multi matrix observables reproduced by the eigenvalue dynamics.
Notice that in this case the agreement is not exact, but rather is realized to the large $N$ limit.
This is what we expect for the generic situation - the AdS$_5\times$S$^5$ case is highly symmetric and the
fact that eigenvalue dynamics reproduces so many observables exactly is a consequence of this symmetry.
We only expect eigenvalue dynamics to reproduce classical gravity, which should emerge from the CFT at $N=\infty$.
\\
\\
Much of our intuition came from thinking about the Gauss graph operators constructed in \cite{Koch:2011hb,deMelloKoch:2012ck}.
It is natural to ask if we can write down wave functions dual to the Gauss graph operators.
The simplest possibility is to consider a Gauss graph operator obtained by exciting a single eigenvalue by $J$ levels, and 
then attaching a total of $K$ $Y$ strings to it.
The extreme simplicity of this case follows because we can write the (normalized) Gauss graph operator in terms of a familiar
Schur polynomial as
\bea
\hat{O}=\sqrt{{J!K!\over (J+K)!}{(N-1)!\over (N+J+K-1)!}}{\rm Tr}\left(Y{\partial\over\partial Z}\right)^K
\chi_{(J+K)}(Z)
\eea
where we have used the notation $(n)$ to denote a Young diagram with a single row of $n$ boxes.
Consider the correlator
\bea
\langle \hat{O}{\rm Tr}(Y^\dagger)^K{\rm Tr}(Z^{\dagger J})\rangle
&=& \langle {\rm Tr}\left({\partial\over\partial Y}\right)^K\hat{O}{\rm Tr}(Z^{\dagger J})\rangle\cr
&=&\sqrt{{J!K!\over (J+K)!}{(N+J+K-1)!\over (N-1)!}}
\eea
This answer is exact, in the free field theory.
In what limit should we compare this answer to eigenvalue dynamics?
Our intuition is coming from the ${1\over 2}$- BPS sector where we know that rows of Schur polynomials correspond to
eigenvalues and we know exactly how to write the corresponding wave function.
If we only want small perturbations of this picture, we should keep $K\ll J$.
In this case we should simplify
\bea
{J!K!\over (J+K)!}&\to& {1\over J^K}\cr
{(N+J+K-1)!\over (N-1)!}&=&{(N+J+K-1)!\over (N+J-1)!}{(N+J-1)!\over (N-1)!}\cr
&\to& (N+J-1)^K{(N+J-1)!\over (N-1)!}
\eea
How should we scale $J$ as we take $N\to\infty$?
The Schur polynomials are a sum over all possible matrix trace structures. 
We want these sums to be dominated by traces with a large number of matrices ($N$ or more) in each trace.
To accomplish this we will scale $J=O(N^{1+\epsilon})$ with $\epsilon>0$.
In this case, at large $N$, we can replace
\bea
{1\over J^K}(N+J-1)^K\to 1
\eea
and hence, the result that should be reproduced by the eigenvalue dynamics is given by
\bea
\langle \hat{O}{\rm Tr}(Y^\dagger)^K{\rm Tr}(Z^{\dagger J})\rangle
=\sqrt{K! {(N+J-1)!\over (N-1)!}}\label{corr}
\eea
In the eigenvalue computation, we will use the wave function of the ground state and
the wave function of the Gauss graph operator ($\Psi_{\rm GG} (\{z_i,\bar z_i,y_i,\bar y_i\})$) to compute the amplitude
\bea
   \int\prod_{i=1}^N dz_i d\bar z_i dy_i d\bar y_i \Psi_{\rm gs}^* (\{z_i,\bar z_i,y_i,\bar y_i\}) (\sum_i\bar y_i)^K
\sum_j \bar z_j^J\Psi_{\rm GG} (\{z_i,\bar z_i,y_i,\bar y_i\}) \label{eamp}
\eea
We expect the amplitude (\ref{eamp}) to reproduce (\ref{corr}).
Our proposal for the wave function corresponding to the above Gauss graph operator is
\bea
\Psi_{GG}(\{z_i,\bar z_i,y_i,\bar y_i\})={\pi^{-N}\over\sqrt{N!}}\epsilon^{a_1 a_2 \cdots a_n} 
{z_{a_1}^0 y_{a_1}^{N-1}\over \sqrt{0!(N-1)!}}\cdots {z_{a_k}^{k-1}y_{a_k}^{N-k}\over\sqrt{(k-1)!(N-k)!}} \cdots\cr
\cdots {z_{a_{N-1}}^{N-2}y_{a_{N-1}}\over\sqrt{(N-2)!1!}}
{z_{a_N}^{J+N-1}y_{a_N}^K\over\sqrt{(J+N-1)!K!}} 
e^{-{1\over 2}\sum_q z_q\bar z_q-{1\over 2}\sum_q y_q\bar y_q}
\eea
The eigenvalue with the largest power of $z$ (i.e. $z_{a_N}$) was the fermion at the very top of the Fermi sea.
It has been excited by $J$ powers of $z$ and $K$ powers of $y$. 
It is now trivial to verify that (\ref{eamp}) does indeed reproduce (\ref{corr}).
\\
\\
Finally, the state with three eigenvalues excited by $J_1>J_2>J_3$ and with $K_1>K_2>K_3$ strings attached to each 
eigenvalue is given by
\bea
\Psi_{GG}(\{z_i,\bar z_i,y_i,\bar y_i\})\!\!\!\! &&\!\!\!\!
={\pi^{-N}\over\sqrt{N!}}\epsilon^{a_1 a_2 \cdots a_n} 
{z_{a_1}^0 y_{a_1}^{N-1}\over \sqrt{0!(N-1)!}}\cdots {z_{a_k}^{k-1}y_{a_k}^{N-k}\over\sqrt{(k-1)!(N-k)!}} \cdots\cr
&&\!\!\!\!\!\!\!\! \cdots
{z_{a_{N-3}}^{N-4}y_{a_{N-3}}^{3}\over\sqrt{(N-4)!3!}}
{z_{a_{N-2}}^{J_3+N-3}y_{a_{N-2}}^{2+K_3}\over\sqrt{(J_3+N-3)!(2+K_3)!}}
{z_{a_{N-1}}^{J_2+N-2}y_{a_{N-1}}^{K_2+1}\over\sqrt{(J_2+N-2)!(K_2+1)!}}\cr
&&\qquad \times {z_{a_N}^{J_1+N-1}y_{a_N}^{K_1}\over\sqrt{(J_1+N-1)!K_1!}} 
e^{-{1\over 2}\sum_q z_q\bar z_q-{1\over 2}\sum_q y_q\bar y_q}
\eea
The generalization to any Gauss graph operator is now clear.

\section{Connection to Supergravity}\label{surfaces}

In this section we would like to explore the possibility that the eigenvalue dynamics of the $SU(2)$ sector has a natural
interpretation in supergravity.
The relevant supergravity solutions have been considered in \cite{Donos:2006iy,Donos:2006ms,Chen:2007du,Lunin:2008tf}.
\\
\\
There are 6 adjoint scalars in the ${\cal N}=4$ super Yang-Mills theory that can be assembled into the following three complex
combinations
\bea
   Z=\phi^1+i\phi^2\qquad    Y=\phi^3+i\phi^4\qquad    X=\phi^5+i\phi^6
\eea 
The operators we consider are constructed using only $Z$ and $Y$ so that they are invariant under the $U(1)$ which 
rotates $\phi^5$ and $\phi^6$.
Further, since our operators are BPS they are built only from the $s$-wave spherical harmonic components of $Y$ and $Z$, so that
they are invariant under the $SO(4)$ symmetry which acts on the $S^3$ of the $R\times S^3$ spacetime on which the CFT is
defined. 
Local supersymmetric geometries with $SO(4)\times U(1)$ isometries have the form\cite{Donos:2006iy,Lunin:2008tf}
\bea
ds_{10}^2=-h^{-2}(dt+\omega)^2+h^2\Big[ {2\over Z+{1\over 2}}\partial_a\bar\partial_b K dz^ad\bar z^b +dy^2\Big]
+y(e^G d\Omega_3^2 + e^{-G}d\psi^2)
\eea
\bea
  d\omega={i\over y}\Big( \partial_a\bar\partial_b \partial_y K dz^ad\bar z^b-\partial_a Zdz^a dy
                  +\bar\partial_a Zd\bar z_a dy\Big)
\eea
Here $z^1$ and $z^2$ is a pair of complex coordinates and $K$ is a Kahler potential which may depend on $y$, $z^a$ 
and $\bar z^a$.
$y^2$ is the product of warp factors for $S^3$ and $S^1$. 
Thus we must be careful and impose the correct boundary conditions at the $y=0$ hypersurface if we are to avoid singularities.
The $y=0$ hypersurface includes the four dimensional space with coordinates given by the $z^a$. 
These boundary conditions require that when the $S^3$ contracts to zero, we need $Z=-{1\over 2}$ and when the
$\psi$-circle collapses we need $Z={1\over 2}$\cite{Donos:2006iy,Lunin:2008tf}.
There is a surface separating these two regions, and hence, defining the supergravity solution.
So far the discussion given closely matches what is found for the ${1\over 2}$-BPS supergravity solutions.
In that case the $y=0$ hypersurface includes a two dimensional space which is similarly divided into two regions, giving
the black droplets on a white plane.
The edges of the droplets are completely arbitrary, which is an important difference from the case we are considering.
The surface defining local supersymmetric geometries with $SO(4)\times U(1)$ isometries is not completely arbitrary - it too has 
to satisfy some additional constraints as spelled out in \cite{Lunin:2008tf}.
It is natural to ask if the surface defining the supergravity solution is visible in the eigenvalue dynamics?
\\
\\
To answer this question we will now review how the surface defining the local supersymmetric geometries with 
$SO(4)\times U(1)$ isometries corresponding to the ${1\over 2}$-BPS LLM geometries is constructed.
According to \cite{Lunin:2008tf}, the boundary condition for these geometries have walls between the two boundary conditions
determined by the equation\footnote{This next equation is (6.35) of \cite{Lunin:2008tf}. We will relate $z^1$ and $z^2$
to $z_i$ (the eigenvalues of $Z$) and $y_i$ (the eigenvalues of $Y$) when we make the correspondence to eigenvalues.}
\bea
z^2\bar z^2 = e^{-2\hat{D}(z^1,\bar z^1)}
\eea
where $\hat{D}(z^1,\bar{z}^1)$ is determined by expanding the function $D$ as follows (it is the $y$ coordinate that 
we set to zero to get the LLM plane)
\bea
D=\log (y)+\hat{D}(z,\bar z)+O(x)
\eea
The function $D$ is determined by the equations
\bea
y\partial_y D={1\over 2}-Z\qquad V=-i(dz\partial_z -d\bar z\partial_{\bar z})D\label{tosolve}
\eea
where $Z(y,z^1,\bar z^1)$ is the function obeying Laplace's equation that determines the LLM solution and $V(y,z^1,\bar z^1)$ 
is the one form appearing in the combination $(dt+V)^2$ in the LLM metric.
\\
\\
Consider an annulus that has an outer edge at radius $M+N$ and an inner edge at a radius $M$.
This solution has (these solutions were constructed in the original LLM paper \cite{Lin:2004nb})
\bea
Z(y,z^1,\bar z^1)=-{1\over 2}\left({|z^1|^2+y^2-M+1\over\sqrt{(|z^1|^2+y^2+M-1)^2-4|z^1|^2 (M-1)}}\right.\cr
\left.+{|z^1|^2+y^2-M-N\over\sqrt{(|z^1|^2+y^2+M+N)^2-4|z^1|^2 (M+N)}}\right)\cr
V(y,z^1,\bar z^1)={d\phi\over 2}\left({|z^1|^2+y^2+M-1\over\sqrt{(|z^1|^2+y^2+M-1)^2-4|z^1|^2 (M-1)}}\right.\cr
\left.+{|z^1|^2+y^2+M+N\over\sqrt{(|z^1|^2+y^2+M+N)^2-4|z^1|^2 (M+N)}}\right)\nonumber
\eea
Evaluating at $y=0$, the second of (\ref{tosolve}) says
\bea
V=-i(dz\partial_z -d\bar z\partial_{\bar z})\hat{D}
\eea
Setting $z^1=re^{-i\phi}$ and assuming that $\hat D$ depends only on $r$ we find
\bea
 r{\partial\hat D\over\partial r}=-{M+N\over r^2-M-N}+{M-1\over r^2-M+1}
\eea
which is solved by
\bea
 \hat D={1\over 2}\log {| z^1\bar z^1 -M+1|\over |z^1\bar z^1 -M-N|}
\eea
Thus, the wall between the two boundary conditions is given by
\bea
  |z^2|^2 ={M+N-1-z^1\bar z^1\over  z^1\bar z^1 -M+1}\label{surface}
\eea

\noindent The same analysis applied to the AdS$_5\times$S$^5$ solution gives
\bea
 |z^1|^2+|z^2|^2=N-1\label{sphere}
\eea
For the pair of geometries described above, we know the wave function in the eigenvalue description.
We will now return to the eigenvalue description and see how these surfaces are related to the eigenvalue wave functions.
\\
\\
At large $N$, since fluctuations are controlled by $1/N^2$, we expect a definite eigenvalue distribution. 
These eigenvalues will trace out a surface specified by the support of the single fermion probability density
\begin{equation}
\rho(z_1,\bar z_1,y_1,\bar y_1)=\int \prod_{i=2}^N dz_id\bar z_i dy_id\bar y_i|\Psi  (\{z_i,\bar z_i,y_i,\bar y_i\})|^2
\label{MMsurf}
\end{equation}
Denote the points lying on this surface using coordinates $z,y$.
\\
\\
Using the wave function $\Psi_{\rm gs} (\{z_i,\bar z_i,y_i,\bar y_i\})$ corresponding to the AdS$_5\times$S$^5$ spacetime, 
the probability density for a single eigenvalue is
\bea
\rho(z,\bar z,y,\bar y)={1\over N \pi^2}\sum_{i=0}^{N-1}
{(z\bar z)^{i}\over i!}{(y\bar y)^{N-i-1}\over (N-i-1)!}e^{-z\bar z-y\bar y}
\eea
As $y$ and $z$ vary, the dominant contribution comes from a term with a specific value for $i$.
When the $i$th term dominates the sum, the value of the eigenvalue coordinate is given by
\bea
{(z\bar z)^{i}\over i!}=1&&\qquad |z|^{2i}=i!\approx i^i\cr
{(y\bar y)^{N-i-1}\over (N-i-1)!}=1&&\qquad |y|^{2(N-i-1)}=(N-i-1)!\approx (N-i-1)^{N-i-1}
\eea
This leads to the following points
\bea
  |z_{(i)}|^2=i\qquad |y_{(i)}|^2=N-i-1\qquad i=0,1,2,...,N-1
\eea
Thus, if we identify the points $z_{(i)},y_{(i)}$ and the supergravity coordinate $z^1,z^2$ as follows
\bea
   z^2=y_{(i)}\qquad z^1=z_{(i)}
\eea
we find
\bea
  |z^1|^2+|z^2|^2=i + (N-i-1)=N-1
\eea
so that the eigenvalues condense on the surface that defines the wall between the two boundary conditions.
\\
\\
Let's now compute the positions of our eigenvalues, using $\Psi_{\rm LLM} (\{z_i,\bar z_i,y_i,\bar y_i\})$.
The probability density for a single eigenvalue is easily obtained by computing the following integral
\bea
\rho(z_1,\bar z_1,y_1,\bar y_1)=\int \prod_{i=2}^N dz_id\bar z_i dy_id\bar y_i|
\Psi_{\rm LLM} (\{z_i,\bar z_i,y_i,\bar y_i\})|^2\cr
={1\over N\pi^2}\sum_{i=0}^{N-1}
{(z_1\bar z_1)^{M+i}\over (M+i)!}{(y_1\bar y_1)^{N-i-1}\over (N-i-1)!}e^{-z_1\bar z_1-y_1\bar y_1}
\eea
Following the analysis we performed above, we find that the complete set of points on the eigenvalue surface is given by
\bea
  |z_{(i)}|^2=(M+i)\qquad |y_{(i)}|^2=N-i-1\qquad i=0,1,2,...,N-1
\eea
Thus, if we identify the points $z_{(i)},y_{(i)}$ and the supergravity coordinate $z^1,z^2$ as follows
\bea
   z^2={y_{(i)}\over\sqrt{|z_{(i)}|^2-M+1}}\qquad z^1=z_{(i)}
\eea
we find that (\ref{surface}) gives
\bea
  {|y^{(i)}|^2\over i+1}={M+N-1- |z^{(i)}|^2\over  |z^{(i)}|^2 -M+1}
\eea
in complete agreement with where our wave function is localized.
This again shows that the eigenvalues are collecting on the surface that defines the wall between the two boundary
conditions.
Although these examples are rather simple, they teach us something important: the map between the eigenvalues 
and the supergravity coordinates depends on the specific geometry we consider.
\\
\\
The fact that eigenvalues condense on the surface that defines the wall between the two boundary conditions is something
that was already anticipated by Berenstein and Cotta in \cite{Berenstein:2007wz}.
The proposal of \cite{Berenstein:2007wz} identifies the support of the eigenvalue distribution with the degeneration locus of 
the three sphere in the full ten dimensional metric.
Our results appear to be in perfect accord with this proposal.

\section{Outlook}\label{outlook}

There are a number of definite conclusions resulting from our study.
One of our key results is that we have found substantial evidence for the proposal that there is a sector of the two matrix 
model that is described (sometimes exactly) by eigenvalue dynamics.
This is rather non-trivial since, as we have already noted, it is simply not true that the two matrices can be simultaneously
diagonalized.
The fact that we have reproduced correlators of operators that involve products of both matrices in a single trace is
convincing evidence that we are reproducing genuine two matrix observables. 
The observables we can reproduce correspond to BPS operators.
In the dual gravity these operators map to supergravity states corresponding to classical geometries.
The local supersymmetric geometries with $SO(4)\times U(1)$ isometries are determined by a surface that defines the
boundary conditions needed to obtain a non-singular supergravity solution.
At large $N$ where we expect classical geometry, the eigenvalues condense on this surface.
In this way the supergravity boundary conditions appear to match the large $N$ eigenvalue description perfectly.
\\
\\
The eigenvalue dynamics appears to provide some sort of a coarse grained description.
Correlators of operators dual to states with a very small energy are not reproduced correctly: for example the energy of states
dual to single traces has to be above some threshold ($N$) before they are correctly reproduced.
For complicated operators with a detailed multi trace structure we would thus expect to get the gross features correct, but we
may miss certain finer details - see the discussion after (\ref{evalschurtwopoint}).
Developing this point of view, perhaps using the ideas outlined in \cite{Balasubramanian:2005mg}, may provide a deeper
understanding of the eigenvalue wave functions.
\\
\\
The eigenvalue description we have developed here is explicit enough that we could formulate the dynamics in terms of the
density of eigenvalues.
This would provide a field theory that has $1/N$ appearing explicitly as a coupling.
It would be very interesting to work out, for example, what the generalization of the Das-Jevicki Hamiltonian\cite{Das:1990kaa}  is. 
\\
\\
The picture of eigenvalue dynamics that we are finding here is almost identical to the 
proposal discussed by Berenstein and his 
collaborators\cite{Berenstein:2005aa,Berenstein:2006yy,Berenstein:2007wz,Berenstein:2008jn,Berenstein:2008me,Berenstein:2010dg,Berenstein:2014pma},
developed using numerical methods and clever heuristic arguments.
The idea of these works is that the eigenvalues represent microscopic degrees of freedom.
At large $N$ one can move to collective degrees of freedom that represent the 10 dimensional geometry of the dual gravitational
description.
This is indeed what we are seeing. 
They have also considered cases with reduced supersymmetry and orbifold 
geometries\cite{Berenstein:2007wi,Berenstein:2007kq,Berenstein:2008eg}.
These are natural examples to consider using the ideas and methods we have developed in this article.
Developing other examples of eigenvalue dynamics will allow us to further test the proposals for wave functions and the large
$N$ distributions of eigenvalues that we have put forward in this article.
\\
\\
An important question that should be tackled is to ask how one could derive (and not guess) the wave functions we have
described.
Progress with this question is likley to give some insights into how it is even possible to have a consistent eigenvalue dynamics.
One would like to know when an eigenvalue description is relevant and to what classes of observables it is applicable.
\\
\\
Another important question is to consider the extension to more matrices, including gauge and fermion degrees of freedom.
The Gauss graph labeling of operators continues to work when we include gauge fields and 
fermions\cite{deMelloKoch:2011vn,Koch:2012sf}, so that our
argument goes through without modification and we again expect that eigenvalue dynamics in these more general settings 
will be an effective approach to compute these more general correlators of BPS operators.
Another important extension is to consider the eigenvalue dynamics, perturbed by off diagonal elements, which should
allow one to start including stringy degrees of freedom.
Can this be done in a controlled systematic fashion?
In this context, the studies carried out in \cite{de Mello Koch:2007uu,de Mello Koch:2007uv,Bekker:2007ea}, will be relevant.

{\vskip 0.5cm}

\noindent
{\it Acknowledgements:}
We would like to thank Joao Rodrigues for interest in this work and for discussions.
This work is based upon research supported by the South African Research Chairs
Initiative of the Department of Science and Technology and National Research Foundation.
Any opinion, findings and conclusions or recommendations expressed in this material
are those of the authors and therefore the NRF and DST do not accept any liability
with regard thereto.


\begin{thebibliography}{} 

\bibitem{'tHooft:1973jz} 
  G.~'t Hooft,
  ``A Planar Diagram Theory for Strong Interactions,''
  Nucl.\ Phys.\ B {\bf 72}, 461 (1974).
  doi:10.1016/0550-3213(74)90154-0

\bibitem{Maldacena:1997re}
  J.~M.~Maldacena,
  ``The large N limit of superconformal field theories and supergravity,''
  Adv.\ Theor.\ Math.\ Phys.\  {\bf 2}, 231 (1998)
  [Int.\ J.\ Theor.\ Phys.\  {\bf 38}, 1113 (1999)]
  [arXiv:hep-th/9711200].

\bibitem{Balasubramanian:2001nh} 
  V.~Balasubramanian, M.~Berkooz, A.~Naqvi and M.~J.~Strassler,
  ``Giant gravitons in conformal field theory,''
  JHEP {\bf 0204}, 034 (2002)
  [hep-th/0107119].

\bibitem{McGreevy:2000cw} 
  J.~McGreevy, L.~Susskind and N.~Toumbas,
  ``Invasion of the giant gravitons from Anti-de Sitter space,''
  JHEP {\bf 0006}, 008 (2000)
  doi:10.1088/1126-6708/2000/06/008
  [hep-th/0003075].

\bibitem{Hashimoto:2000zp} 
  A.~Hashimoto, S.~Hirano and N.~Itzhaki,
  ``Large branes in AdS and their field theory dual,''
  JHEP {\bf 0008}, 051 (2000)
  doi:10.1088/1126-6708/2000/08/051
  [hep-th/0008016].

\bibitem{Grisaru:2000zn} 
  M.~T.~Grisaru, R.~C.~Myers and O.~Tafjord,
  ``SUSY and goliath,''
  JHEP {\bf 0008}, 040 (2000)
  doi:10.1088/1126-6708/2000/08/040
  [hep-th/0008015].

\bibitem{Lin:2004nb} 
  H.~Lin, O.~Lunin and J.~M.~Maldacena,
  ``Bubbling AdS space and 1/2 BPS geometries,''
  JHEP {\bf 0410}, 025 (2004)
  [hep-th/0409174].

\bibitem{Corley:2001zk} 
  S.~Corley, A.~Jevicki and S.~Ramgoolam,
  ``Exact correlators of giant gravitons from dual N=4 SYM theory,''
  Adv.\ Theor.\ Math.\ Phys.\  {\bf 5}, 809 (2002)
  [hep-th/0111222].

\bibitem{Berenstein:2004kk} 
  D.~Berenstein,
  ``A Toy model for the AdS / CFT correspondence,''
  JHEP {\bf 0407}, 018 (2004)
  doi:10.1088/1126-6708/2004/07/018
  [hep-th/0403110].

\bibitem{BIPZ}
E.~Brezin, C.~Itzykson, G.~Parisi and J.~B.~Zuber,
  ``Planar Diagrams,''
  Commun.\ Math.\ Phys.\  {\bf 59}, 35 (1978).
  doi:10.1007/BF01614153

\bibitem{Ginibre:1965zz} 
  J.~Ginibre,
  ``Statistical Ensembles of Complex, Quaternion and Real Matrices,''
  J.\ Math.\ Phys.\  {\bf 6}, 440 (1965).
  doi:10.1063/1.1704292

\bibitem{Jevicki:1979mb} 
  A.~Jevicki and B.~Sakita,
  ``The Quantum Collective Field Method and Its Application to the Planar Limit,''
  Nucl.\ Phys.\ B {\bf 165}, 511 (1980).
  doi:10.1016/0550-3213(80)90046-2

\bibitem{Jevicki:1980zg} 
  A.~Jevicki and B.~Sakita,
  ``Collective Field Approach to the Large $N$ Limit: Euclidean Field Theories,''
  Nucl.\ Phys.\ B {\bf 185}, 89 (1981).
  doi:10.1016/0550-3213(81)90365-5

\bibitem{Demeterfi:1991tz} 
  K.~Demeterfi, A.~Jevicki and J.~P.~Rodrigues,
  ``Scattering amplitudes and loop corrections in collective string field theory,''
  Nucl.\ Phys.\ B {\bf 362}, 173 (1991).
  doi:10.1016/0550-3213(91)90561-B

\bibitem{Demeterfi:1991nw} 
  K.~Demeterfi, A.~Jevicki and J.~P.~Rodrigues,
  ``Scattering amplitudes and loop corrections in collective string field theory. 2.,''
  Nucl.\ Phys.\ B {\bf 365}, 499 (1991).
  doi:10.1016/S0550-3213(05)80030-6

\bibitem{Demeterfi:1991cw} 
  K.~Demeterfi, A.~Jevicki and J.~P.~Rodrigues,
  ``Perturbative results of collective string field theory,''
  Mod.\ Phys.\ Lett.\ A {\bf 6}, 3199 (1991).
  doi:10.1142/S0217732391003699

\bibitem{Donos:2005vm} 
  A.~Donos, A.~Jevicki and J.~P.~Rodrigues,
  ``Matrix model maps in AdS/CFT,''
  Phys.\ Rev.\ D {\bf 72}, 125009 (2005)
  doi:10.1103/PhysRevD.72.125009
  [hep-th/0507124].

\bibitem{Masuku:2009qf} 
  M.~Masuku and J.~P.~Rodrigues,
  ``Laplacians in polar matrix coordinates and radial fermionization in higher dimensions,''
  J.\ Math.\ Phys.\  {\bf 52}, 032302 (2011)
  doi:10.1063/1.3553456
  [arXiv:0911.2846 [hep-th]].

\bibitem{Kimura:2009ur} 
  Y.~Kimura, S.~Ramgoolam and D.~Turton,
  ``Free particles from Brauer algebras in complex matrix models,''
  JHEP {\bf 1005}, 052 (2010)
  doi:10.1007/JHEP05(2010)052
  [arXiv:0911.4408 [hep-th]].

\bibitem{Masuku:2011pm} 
  M.~Masuku and J.~P.~Rodrigues,
  ``How universal is the Wigner distribution?,''
  J.\ Phys.\ A {\bf 45}, 085201 (2012)
  doi:10.1088/1751-8113/45/8/085201
  [arXiv:1107.3681 [hep-th]].

\bibitem{Masuku:2014wxa} 
  M.~Masuku, M.~Mulokwe and J.~P.~Rodrigues,
  ``Large N Matrix Hyperspheres and the Gauge-Gravity Correspondence,''
  JHEP {\bf 1512}, 035 (2015)
  doi:10.1007/JHEP12(2015)035
  [arXiv:1411.5786 [hep-th]].

\bibitem{Masuku:2015vta} 
  M.~Masuku and J.~P.~Rodrigues,
  ``De Alfaro, Fubini and Furlan from multi Matrix Systems,''
  JHEP {\bf 1512}, 175 (2015)
  doi:10.1007/JHEP12(2015)175
  [arXiv:1509.06719 [hep-th]].

\bibitem{Kristjansen:2002bb} 
  C.~Kristjansen, J.~Plefka, G.~W.~Semenoff and M.~Staudacher,
  ``A New double scaling limit of N=4 superYang-Mills theory and PP wave strings,''
  Nucl.\ Phys.\ B {\bf 643}, 3 (2002)
  doi:10.1016/S0550-3213(02)00749-6
  [hep-th/0205033].

\bibitem{Takayama:2005yq} 
  Y.~Takayama and A.~Tsuchiya,
  ``Complex matrix model and fermion phase space for bubbling AdS geometries,''
  JHEP {\bf 0510}, 004 (2005)
  doi:10.1088/1126-6708/2005/10/004
  [hep-th/0507070].

\bibitem{Balasubramanian:2004nb} 
  V.~Balasubramanian, D.~Berenstein, B.~Feng and M.~x.~Huang,
  ``D-branes in Yang-Mills theory and emergent gauge symmetry,''
  JHEP {\bf 0503}, 006 (2005)
  [hep-th/0411205].

\bibitem{Bhattacharyya:2008rb} 
  R.~Bhattacharyya, S.~Collins and R.~d.~M.~Koch,
 ``Exact Multi-Matrix Correlators,''
  JHEP {\bf 0803}, 044 (2008)
  [arXiv:0801.2061 [hep-th]].

\bibitem{Bhattacharyya:2008xy} 
  R.~Bhattacharyya, R.~de Mello Koch and M.~Stephanou,
  ``Exact Multi-Restricted Schur Polynomial Correlators,''
  JHEP {\bf 0806}, 101 (2008)
  [arXiv:0805.3025 [hep-th]].

\bibitem{DeComarmond:2010ie} 
  V.~De Comarmond, R.~de Mello Koch and K.~Jefferies,
  ``Surprisingly Simple Spectra,''
  JHEP {\bf 1102}, 006 (2011)
  [arXiv:1012.3884 [hep-th]].

\bibitem{Koch:2011hb} 
  R.~d.~M.~Koch, M.~Dessein, D.~Giataganas and C.~Mathwin,
  ``Giant Graviton Oscillators,''
  JHEP {\bf 1110}, 009 (2011)
  [arXiv:1108.2761 [hep-th]].

\bibitem{deMelloKoch:2012ck} 
  R.~de Mello Koch and S.~Ramgoolam,
  ``A double coset ansatz for integrability in AdS/CFT,''
  JHEP {\bf 1206}, 083 (2012)
  doi:10.1007/JHEP06(2012)083
  [arXiv:1204.2153 [hep-th]].

\bibitem{Berenstein:2005aa} 
  D.~Berenstein,
  ``Large N BPS states and emergent quantum gravity,''
  JHEP {\bf 0601}, 125 (2006)
  doi:10.1088/1126-6708/2006/01/125
  [hep-th/0507203].

\bibitem{Berenstein:2006yy} 
  D.~Berenstein and R.~Cotta,
  ``Aspects of emergent geometry in the AdS/CFT context,''
  Phys.\ Rev.\ D {\bf 74}, 026006 (2006)
  doi:10.1103/PhysRevD.74.026006
  [hep-th/0605220].

\bibitem{Berenstein:2007wz} 
  D.~Berenstein and R.~Cotta,
  ``A Monte-Carlo study of the AdS/CFT correspondence: An Exploration of quantum gravity effects,''
  JHEP {\bf 0704}, 071 (2007)
  doi:10.1088/1126-6708/2007/04/071
  [hep-th/0702090].

\bibitem{Berenstein:2008jn} 
  D.~Berenstein, R.~Cotta and R.~Leonardi,
  ``Numerical tests of AdS/CFT at strong coupling,''
  Phys.\ Rev.\ D {\bf 78}, 025008 (2008)
  doi:10.1103/PhysRevD.78.025008
  [arXiv:0801.2739 [hep-th]].

\bibitem{Berenstein:2008me} 
  D.~Berenstein,
  ``A Strong coupling expansion for N=4 SYM theory and other SCFT's,''
  Int.\ J.\ Mod.\ Phys.\ A {\bf 23}, 2143 (2008)
  doi:10.1142/S0217751X08040688
  [arXiv:0804.0383 [hep-th]].

\bibitem{Berenstein:2010dg} 
  D.~Berenstein and Y.~Nakada,
  ``The Shape of Emergent Quantum Geometry from an N=4 SYM Minisuperspace Approximation,''
  arXiv:1001.4509 [hep-th].

\bibitem{Berenstein:2014pma} 
  D.~Berenstein,
  ``Sketches of emergent geometry in the gauge/gravity duality,''
  Fortsch.\ Phys.\  {\bf 62}, 776 (2014)
  [arXiv:1404.7052 [hep-th]].

\bibitem{Balasubramanian:2005mg} 
  V.~Balasubramanian, J.~de Boer, V.~Jejjala and J.~Simon,
  ``The Library of Babel: On the origin of gravitational thermodynamics,''
  JHEP {\bf 0512}, 006 (2005)
  doi:10.1088/1126-6708/2005/12/006
  [hep-th/0508023].

\bibitem{Vazquez:2006id}
  S.~E.~Vazquez,
  ``Reconstructing 1/2 BPS space-time metrics from matrix models and spin
  chains,''
  Phys.\ Rev.\  D {\bf 75}, 125012 (2007)
  [arXiv:hep-th/0612014].

\bibitem{Chen:2007gh}
  H.~Y.~Chen, D.~H.~Correa and G.~A.~Silva,
  ``Geometry and topology of bubble solutions from gauge theory,''
  Phys.\ Rev.\  D {\bf 76}, 026003 (2007)
  [arXiv:hep-th/0703068].

\bibitem{Koch:2008ah} 
  R.~de Mello Koch,
  ``Geometries from Young Diagrams,''
  JHEP {\bf 0811}, 061 (2008)
  doi:10.1088/1126-6708/2008/11/061
  [arXiv:0806.0685 [hep-th]].

\bibitem{Lee:1998bxa} 
  S.~Lee, S.~Minwalla, M.~Rangamani and N.~Seiberg,
  ``Three point functions of chiral operators in D = 4, N=4 SYM at large N,''
  Adv.\ Theor.\ Math.\ Phys.\  {\bf 2}, 697 (1998)
  [hep-th/9806074].

\bibitem{Corley:2002mj} 
  S.~Corley and S.~Ramgoolam,
  ``Finite factorization equations and sum rules for BPS correlators in N=4 SYM theory,''
  Nucl.\ Phys.\ B {\bf 641}, 131 (2002)
  doi:10.1016/S0550-3213(02)00573-4
  [hep-th/0205221].

\bibitem{Myers:2001aq} 
  R.~C.~Myers and O.~Tafjord,
  ``Superstars and giant gravitons,''
  JHEP {\bf 0111}, 009 (2001)
  doi:10.1088/1126-6708/2001/11/009
  [hep-th/0109127].

\bibitem{Donos:2006iy} 
  A.~Donos,
  ``A Description of 1/4 BPS configurations in minimal type IIB SUGRA,''
  Phys.\ Rev.\ D {\bf 75}, 025010 (2007)
  doi:10.1103/PhysRevD.75.025010
  [hep-th/0606199].

\bibitem{Donos:2006ms} 
  A.~Donos,
  ``BPS states in type IIB SUGRA with SO(4) x SO(2)(gauged) symmetry,''
  JHEP {\bf 0705}, 072 (2007)
  doi:10.1088/1126-6708/2007/05/072
  [hep-th/0610259].

\bibitem{Chen:2007du} 
  B.~Chen, S.~Cremonini, A.~Donos, F.~L.~Lin, H.~Lin, J.~T.~Liu, D.~Vaman and W.~Y.~Wen,
  ``Bubbling AdS and droplet descriptions of BPS geometries in IIB supergravity,''
  JHEP {\bf 0710}, 003 (2007)
  doi:10.1088/1126-6708/2007/10/003
  [arXiv:0704.2233 [hep-th]].

\bibitem{Lunin:2008tf} 
  O.~Lunin,
  ``Brane webs and 1/4-BPS geometries,''
  JHEP {\bf 0809}, 028 (2008)
  doi:10.1088/1126-6708/2008/09/028
  [arXiv:0802.0735 [hep-th]].

\bibitem{Das:1990kaa} 
  S.~R.~Das and A.~Jevicki,
  ``String Field Theory and Physical Interpretation of $D=1$ Strings,''
  Mod.\ Phys.\ Lett.\ A {\bf 5}, 1639 (1990).
  doi:10.1142/S0217732390001888

\bibitem{Berenstein:2007wi} 
  D.~Berenstein,
  ``Strings on conifolds from strong coupling dynamics, part I,''
  JHEP {\bf 0804}, 002 (2008)
  doi:10.1088/1126-6708/2008/04/002
  [arXiv:0710.2086 [hep-th]].

\bibitem{Berenstein:2007kq} 
  D.~E.~Berenstein and S.~A.~Hartnoll,
  ``Strings on conifolds from strong coupling dynamics: Quantitative results,''
  JHEP {\bf 0803}, 072 (2008)
  doi:10.1088/1126-6708/2008/03/072
  [arXiv:0711.3026 [hep-th]].

\bibitem{Berenstein:2008eg} 
  D.~E.~Berenstein, M.~Hanada and S.~A.~Hartnoll,
  ``Multi-matrix models and emergent geometry,''
  JHEP {\bf 0902}, 010 (2009)
  doi:10.1088/1126-6708/2009/02/010
  [arXiv:0805.4658 [hep-th]].

\bibitem{deMelloKoch:2011vn} 
  R.~de Mello Koch, P.~Diaz and H.~Soltanpanahi,
  ``Non-planar Anomalous Dimensions in the sl(2) Sector,''
  Phys.\ Lett.\ B {\bf 713}, 509 (2012)
  doi:10.1016/j.physletb.2012.06.057
  [arXiv:1111.6385 [hep-th]].

\bibitem{Koch:2012sf} 
  R.~de Mello Koch, P.~Diaz and N.~Nokwara,
  ``Restricted Schur Polynomials for Fermions and integrability in the su(2|3) sector,''
  JHEP {\bf 1303}, 173 (2013)
  doi:10.1007/JHEP03(2013)173
  [arXiv:1212.5935 [hep-th]].

\bibitem{de Mello Koch:2007uu} 
  R.~de Mello Koch, J.~Smolic and M.~Smolic,
  ``Giant Gravitons - with Strings Attached (I),''
  JHEP {\bf 0706}, 074 (2007)
  [hep-th/0701066].

\bibitem{de Mello Koch:2007uv} 
  R.~de Mello Koch, J.~Smolic and M.~Smolic,
  ``Giant Gravitons - with Strings Attached (II),''
  JHEP {\bf 0709}, 049 (2007)
  [hep-th/0701067].

\bibitem{Bekker:2007ea} 
  D.~Bekker, R.~de Mello Koch and M.~Stephanou,
  ``Giant Gravitons - with Strings Attached. III.,''
  JHEP {\bf 0802}, 029 (2008)
  [arXiv:0710.5372 [hep-th]].

\end{thebibliography}
\end{document}